\documentstyle[aps, preprint,eqsecnum, psfrag]{revtex}

\tightenlines

\begin{document}

\draft

\title{A Classification of 2d Random Dirac Fermions.}
\author{ 
{ Denis Bernard$^\spadesuit$}\footnote{Member of the C. N. R. S.}   
and  Andr\'e  LeClair$^{\clubsuit}$}

\address{
$^\spadesuit$ Service de Physique Th\'eorique de Saclay
\footnote{Laboratoire de la Direction des Sciences de la Mati\`ere 
du Commisariat \`a l'Energie Atomique, URA 2306 du C. N. R. S.}
F-91191, Gif-sur-Yvette, France. }  
\address{$^\clubsuit$ Newman Laboratory, Cornell University,   
Ithaca, NY 14853.}

\date{September 2001}
\maketitle

\begin{abstract}
We present a detailed classification of random Dirac hamiltonians
in two spatial dimensions  based on the implementation of  
discrete symmetries.   Our  classification 
is slightly finer than that of random matrices, and contains 
thirteen classes.
We also extend this classification to non-hermitian hamiltonians 
with and without Dirac structure. 
\end{abstract}

\vskip 0.2cm

%
%
\def\debut{ \begin{eqnarray} }
\def\fin{ \end{eqnarray} }
\def\non{ \nonumber }
%
%
\def\betaf{$\beta~$}
\def\oti{{\otimes}}
\def\bra#1{{\langle #1 |  }}
\def\lb{ \left[ }
\def\rb{ \right]  }
\def\tilde{\widetilde}
\def\bar{\overline}
\def\hat{\widehat}
\def\*{\star}
\def\[{\left[}
\def\]{\right]}
\def\({\left(}          \def\BL{\Bigr(}
\def\){\right)}         \def\BR{\Bigr)}
        \def\BBL{\lb}
        \def\BBR{\rb}
%
%
\def\gp{g_+}
\def\betab{{\bar{\beta}}}
\def\zb{{\bar{z} }}
\def\zbar{{\bar{z} }}
\def\frac#1#2{{#1 \over #2}}
\def\inv#1{{1 \over #1}}
\def\half{{1 \over 2}}
\def\d{\partial}
\def\der#1{{\partial \over \partial #1}}
\def\dd#1#2{{\partial #1 \over \partial #2}}
\def\vev#1{\langle #1 \rangle}
\def\ket#1{ | #1 \rangle}
\def\rvac{\hbox{$\vert 0\rangle$}}
\def\lvac{\hbox{$\langle 0 \vert $}}
\def\2pi{\hbox{$2\pi i$}}
\def\e#1{{\rm e}^{^{\textstyle #1}}}
\def\grad#1{\,\nabla\!_{{#1}}\,}
\def\dsl{\raise.15ex\hbox{/}\kern-.57em\partial}
\def\Dsl{\,\raise.15ex\hbox{/}\mkern-.13.5mu D}
%
%
\def\dx{\frac{d^2 x}{2\pi}}
\def\th{\theta}         \def\Th{\Theta}
\def\ga{\gamma}         \def\Ga{\Gamma}
\def\be{\beta}
\def\al{\alpha}
\def\ep{\epsilon}
\def\vep{\varepsilon}
\def\la{\lambda}        \def\La{\Lambda}
\def\de{\delta}         \def\De{\Delta}
\def\om{\omega}         \def\Om{\Omega}
\def\sig{\sigma}        \def\Sig{\Sigma}
\def\vphi{\varphi}
%
%
\def\CA{{\cal A}}       \def\CB{{\cal B}}       \def\CC{{\cal C}}
\def\CD{{\cal D}}       \def\CE{{\cal E}}       \def\CF{{\cal F}}
\def\CG{{\cal G}}       \def\CH{{\cal H}}       \def\CI{{\cal J}}
\def\CJ{{\cal J}}       \def\CK{{\cal K}}       \def\CL{{\cal L}}
\def\CM{{\cal M}}       \def\CN{{\cal N}}       \def\CO{{\cal O}}
\def\CP{{\cal P}}       \def\CQ{{\cal Q}}       \def\CR{{\cal R}}
\def\CS{{\cal S}}       \def\CT{{\cal T}}       \def\CU{{\cal U}}
\def\CV{{\cal V}}       \def\CW{{\cal W}}       \def\CX{{\cal X}}
\def\CY{{\cal Y}}       \def\CZ{{\cal Z}}
\def\rvac{\hbox{$\vert 0\rangle$}}
\def\lvac{\hbox{$\langle 0 \vert $}}
\def\comm#1#2{ \BBL\ #1\ ,\ #2 \BBR }
\def\2pi{\hbox{$2\pi i$}}
\def\e#1{{\rm e}^{^{\textstyle #1}}}
\def\grad#1{\,\nabla\!_{{#1}}\,}
\def\dsl{\raise.15ex\hbox{/}\kern-.57em\partial}
\def\Dsl{\,\raise.15ex\hbox{/}\mkern-.13.5mu D}
%
%
%
\font\numbers=cmss12
\font\upright=cmu10 scaled\magstep1
\def\stroke{\vrule height8pt width0.4pt depth-0.1pt}
\def\topfleck{\vrule height8pt width0.5pt depth-5.9pt}
\def\botfleck{\vrule height2pt width0.5pt depth0.1pt}
\def\Zmath{\vcenter{\hbox{\numbers\rlap{\rlap{Z}\kern
0.8pt\topfleck}\kern 2.2pt
                   \rlap Z\kern 6pt\botfleck\kern 1pt}}}
\def\Qmath{\vcenter{\hbox{\upright\rlap{\rlap{Q}\kern
                   3.8pt\stroke}\phantom{Q}}}}
\def\Nmath{\vcenter{\hbox{\upright\rlap{I}\kern 1.7pt N}}}
\def\Cmath{\vcenter{\hbox{\upright\rlap{\rlap{C}\kern
                   3.8pt\stroke}\phantom{C}}}}
\def\Rmath{\vcenter{\hbox{\upright\rlap{I}\kern 1.7pt R}}}
\def\Z{\ifmmode\Zmath\else$\Zmath$\fi}
\def\Q{\ifmmode\Qmath\else$\Qmath$\fi}
\def\N{\ifmmode\Nmath\else$\Nmath$\fi}
\def\C{\ifmmode\Cmath\else$\Cmath$\fi}
\def\R{\ifmmode\Rmath\else$\Rmath$\fi}







\def\zb{ {\bar z} }

\section{Introduction and results.}

Recently, there is an increasing interest in 
two dimensional localization problems whose  behavior  
differs  from  generic Anderson localization \cite{Anderson}.
To mention but a few examples,  there are investigations   
of the quantum Hall  transition 
\cite{Pruisken,ChalkCodd,ZirnQHE,LFSG}, 
of quasi-particle localization in systems with degenerate 
Fermi surfaces \cite{Fradkin}  and  dirty superconductors
\cite{NTW,SenFis,BocSerZ,AlSiZirn}, and  studies of   
hopping models on bipartite lattices \cite{Gade,HWK}.

Universality classes of localization/delocalization 
transitions  
depend largely on their discrete symmetries.  
Thus the Wigner-Dyson
classification of random hermitian matrices
\cite{Dyson}\cite{Mehta} plays a significant role.
Localization in superconductors led Altland and Zirnbauer
to  significantly extend  the Wigner-Dyson  classification by 
incorporating 
particle-hole and chirality symmetries of the matrices \cite{Zirn}.

Most of the localization problems mentioned above may be formulated  
as spectral problems for Dirac-like hamiltonians in two spacial dimensions,
and many  of them differ from generic Anderson localization  in that 
they exhibit a singular density of states at the critical point.   

In this note, we present a somewhat detailed classification
of such  Dirac hamiltonians in two dimensions.  
This classification  potentially differ from that of  
random matrices because in the latter no structure is imposed
on  the matrices --- and matrices differing by unitary similarity
transformations are treated as equivalent --- whereas in the
former the hamiltonians a priori possess a Dirac form.
This may have two opposite effects: either some random matrix classes
may not be realized by Dirac operators, or Dirac hamiltonians 
belonging to the same random matrix class may not be equivalent
if the unitary similarity transformation relating them
does not preserve the imposed Dirac structure.   Although some models of
random Dirac fermions have  already  been identified with  
Altland-Zirnbauer classes,  the correspondences have not been fully
established and  one  motivation of our classification 
is a more complete dictionary.

We consider Dirac hamiltonians
$H = ( \tau_x p_x + \tau_y p_y )/2 +\vec{\tau}\cdot \vec{W} + W_0$
where $\vec{\tau}$ are Pauli matrices\footnote{
Pauli matrices will be denoted $\vec{\tau}$ or $\vec{\sigma}$
depending on which space they are acting. Our convention is
$\sigma_z=\pmatrix{1&0\cr0&-1\cr},\ 
\sigma_x=\pmatrix{0&1\cr1&0\cr},
\sigma_y=\pmatrix{0&i\cr-i&0\cr}$.}, 
$p_{x,y} = -i \d_{x,y}$, and $W_0$ and $\vec{W}$ are generalized 
masses or  potentials, and  are matrices acting on an isopsin sector.
Introducing complex coordinates, $z=x+iy$, $\zb = x-iy$,
and $\d_z = (\d_x - i \d_y)/2 $, $\d_\zb =( \d_x + i\d_y)/2 $, after
a unitary transformation 
one generally obtains the following $2\times 2$ block structure:
\debut
H = \left(\matrix{ V_+ + V_- & -i \d_\zb + A_\zb 
         \cr  -i \d_z + A_z & V_+-V_- \cr} \right) 
\label{hamilD}
\fin
Here $A_z$, $A_\zb$ and $V_\pm$ are random matrices depending on
the spacial coordinates $x,y$,   and belonging to some statistical ensemble.

As usual, the classes are sets of hamiltonians with specific
transformation properties under some discrete symmetries.
For Dirac hamiltonians  (\ref{hamilD}),   the  simplest symmetries  are
chiral, particle-hole,  and  time-reversal symmetry,    
which  relate the hamiltonian $H$   to  $-H$,  
its  transpose $H^T$ and its complex conjugate $H^*$ respectively.   
We demand that these transformations are implemented by
unitary transformations and that their actions on the  hamiltonian 
square to one. They also should preserve the form (\ref{hamilD})
of the Dirac hamiltonian.
Hence we consider the following transformations:
\debut
{\rm P\ sym.}&:&\quad H = - P\, H \, P^{-1},\quad 
P=\pmatrix{\ga & 0\cr 0& -\ga\cr},\ PP^{\dag}=1,\ P^2=1
\label{Psym}\\
{\rm C\ sym.}&:&\quad H = \ep_c\, C\, H^T \, C^{-1},\quad
C=\pmatrix{0& \sigma \cr  -\ep_c\sigma &0 \cr},\ CC^{\dag}=1,\  C^T=\pm C
\label{Csym}\\
{\rm K\ sym.}&:&\quad H = \ep_k\, K\, H^* \, K^{-1},\quad
C=\pmatrix{0& \kappa \cr  -\ep_k\kappa &0 \cr},\ KK^{\dag}=1,\  K^T=\pm K
\label{Ksym}
\fin
where $\ep_c = \pm 1$. 
Type $P$ symmetries are commonly referred to as chirality symmetries, 
$C$ expresses a particle-hole symmetry, and $K$ time-reversal symmetry. 
For hermitian hamiltonians,  since $H^T=H^*$, $C$ and $K$ symmetries are 
identical and we will only talk about $C$ symmetry, where $\ep_c=+1$ will
be interpreted as time-reversal symmetry and $\ep_c = -1$ will be referred to
as particle-hole symmetry.    

We found thirteen distinct classes of hermitian Dirac hamiltonians,
listed in eqs.(\ref{class0}--\ref{class9}).
This classification, which is presented in
Section II, is slightly finer than that of 
random hermitian matrices \cite{Zirn}.
See Table 1 for a comparison. (The numbering of the classes
has no special meaning.)  
The essential difference between these two classifications 
is a doubling of the chiral classes. 
This arises from differences in the  notion of equivalent hamiltonians.

\begin{center}
\begin{tabular}{|c||c|c|c|c|c|}
\hline
 ~~ & Random matrix & Time rev. & Part.-hole &
 Chirality & Sym. \\
 ~~~ & class  & inv. & symmetry & & group \\
\hline
\hline
class {\bf 0}~~ & ${\bf A}$=GUE & no & no & no & $U(n)$ \\
\hline 
class ${\bf 1}$~~ & ${\bf A_{III}}$=chiral GUE & no & no & yes & $U(n)$ \\
\hline 
class ${\bf 2}$~~ & ${\bf A_{III}}$=chiral GUE 
& no & no & yes & $U(n)\times U(n)$\\
\hline 
class ${\bf 3}_+$ & ${\bf A_{II}}$=G0E & yes  & no & no & $O(n)$ \\
\hline
class ${\bf 3}_-$ & ${\bf D}$  & no  & yes & no & $O(n)$ \\
\hline
class ${\bf 4}_+$ & ${\bf A_{I}}$=GSE  & yes & no & no & $Sp(2n)$ \\
\hline
class ${\bf 4}_-$ & ${\bf C}$ & no & yes & no & $Sp(2n)$ \\
\hline
class ${\bf 5}$~~ & ${\bf D_{III}}$=chiral GOE
 & yes & yes & yes & $O(n)$ \\
\hline
class ${\bf 6}$~~ & ${\bf C_{I}}$
=chiral GSE & yes & yes & yes & $Sp(2n)$ \\
\hline
class ${\bf 7}~~$ & ${\bf D_{III}}$=chiral GOE 
& yes  & yes & yes & $O(n)\times O(n)$ \\
\hline
class ${\bf 8}~~$ & ${\bf C_{I}}$=chiral GSE 
& yes & yes  & yes & $Sp(2n)\times Sp(2n)$ \\
\hline
class ${\bf 9}_+$ & ${\bf D_{I}}$ & yes & yes  & yes & $U(n)$ \\
\hline
class ${\bf 9}_-$ & ${\bf C_{II}}$ & yes & yes & yes & $U(n)$ \\
\hline
\end{tabular}
\end{center}

Let us mention a few well-known  
realizations of the classes of Table 1.
Classes ${\bf 0}$ (GUE), ${\bf 3}_+$ (GOE) and ${\bf 4_+}$ (GSE)
are the usual Wigner-Dyson classes.
The $U(1)$ model of class ${\bf 0}$ was introduced 
in \cite{LFSG} in connection with the quantum Hall transition.
The $U(n)$ case of class ${\bf 0}$ appeared in \cite{Fradkin} 
for describing systems with degenerate Fermi points.
The chiral classes ${\bf 1}$ (chGUE), ${\bf 5}$ (chGOE) 
and ${\bf 6}$ (chGSE) are realized by Dirac operators 
coupled only to random gauge potentials \cite{Verbaa}. 
The $U(n)$ model of class ${\bf 1}$ 
was applied to dirty d-wave superconductors 
in  \cite{NTW}. The pure $U(1)$ random gauge potential
is still not completely solved due to some recently recognized
non-perturbative effects\cite{CCFGM,LeD,Husetal};  
the non-abelian cases have been solved
by various methods \cite{NTW,Cargese,BL}.
The chiral class ${\bf 2}$ (chGUE) is realized by the Gade-Wegner 
hopping models \cite{Gade}.  (See \cite{Hutsa} for a recent
numerical study.)   The class ${\bf 3}_-$ (D) \cite{classD} 
and class ${\bf 4}_-$ (C) \cite{classC}
appeared in the context of dirty superconductors with broken
time reversal symmetry. 
Numerical analysis of the $sp(2)$ model of class ${\bf 4}_-$ (C),
the so-called spin quantum Hall effect, was performed in
\cite{spinqhe}
and exact results were obtained by mapping it to percolation \cite{GLR}.

Since each of the ten classes of random matrices has been associated
with a sigma-model on a symmetric space,  an interesting open question
concerns how the sigma-models can incorporate the finer classification
described in this paper.   In particular the renormalizable effective
field theories described in section II  can have 1, 2, ... up to 10
couplings  
whereas sigma models on symmetric spaces generally have a single coupling
and   possibly an   additional coupling coming from a topological
$\theta$-term  or Wess-Zumino term.

Non-hermitian random hamiltonians have recently been used in the
description of various phenomena, see e.g.\cite{Nelson}.
In Section III we  extend our classification to non-hermitian
hamiltonians. 
In addition to type $P, C$ and $K$ symmetries, 
we may  consider a $Q$ symmetry relating $H$ to its adjoint: 
\begin{equation}
{\rm Q\ sym.}:\quad H = \ep_q\, Q\, H^{\dag} \, Q^{-1},\quad 
Q=\pmatrix{\xi & 0\cr 0& \ep_q\xi\cr},\ QQ^{\dag}=1,\ Q^2=1
\label{Qsym}
\end{equation}
Imposing these symmetries selects reality conditions on the potentials
$A_z$, $A_\zb$ and $V_\pm$.
Of course, the type $Q$ symmetry with $\ep_q=1$ and $\xi=1$ simply
means that $H$ is hermitian.  From
 any non hermitian Dirac operator one may naturally define 
a hermitian one, denoted $\CH$, by doubling 
Hilbert space on which it acts.  (See eq.(\ref{double}).)
The latter hamiltonian is then a representative of the chiral
class we have indexed as class ${\bf 2}$ (chGUE). 
However, as we explain,  the classification of the non-hermitian
Dirac operators $H$ is finer, and more involved, than 
that of their doubled companions $\CH$.  We find a total of 
$87$ universality classes.   

As a byproduct of our analysis we are easily able to classify
non-hermitian random matrices without a Dirac structure.   
This leads to 43 classes.

\section{Classification of hermitian Dirac hamiltonians.} 
We first consider hermitian hamiltonians which
requires $V_\pm^{\dag}=V_\pm$ and $A_z^{\dag}=A_\zb$.
Let us define a ``minimal class'' as a 
class of Dirac hamiltonians which cannot 
be simultaneously block diagonalized.
For such hamiltonians there exists no fixed unitary matrix
$S$ that commutes with the hamiltonian,  
$H=S\,H\,S^{-1}$, and preserves the Dirac structure, which 
requires  
$S={\rm diag}(s,s)$.
The existence of an integral of motion, such as spin, implies
such an $S$ and the resulting hamiltonian is thus not minimal according
to our definition.   Rather our ensembles apply to each block
with fixed quantum numbers of the integrals of motion.

\subsection{Compatible symmetries.}
Only type $P$ and type $C$ symmetries are relevant for hermitian hamiltonians.
We first need to classify the compatible operators $P$ and $C$,
or preferably the compatible  $\ga$ and $\sigma$.
It is important to bear in mind that what is  meaningful is
the group generated by these symmetries.  For instance, 
if the hamiltonian possesses both a
$P$ and a $C$ symmetry,  then it automatically has another
$C$-type symmetry $C'$: 
\begin{equation}
\label{PC}
H = \ep'_c C' \,  H^T  \, C'^{-1},  ~~~~~ C' = PC, ~~~\ep'_c = -\ep_c
\end{equation}
For hermitian hamiltonians, since $C'$ can be interpreted as
a time-reversal (particle-hole)  symmetry if $\ep_c = -1$ ($\ep_c = +1$), 
the  classes with both a $P$ and $C$ symmetry thus automatically
have chirality,  particle-hole and time-reversal symmetry. 
 
The operators $P$ and $C$ are defined up to dilatations by 
scalars and up to unitary changes of basis, $H\to U\,H\,U^{\dag}$,
which 
preserve the form of the Dirac hamiltonians. 
This requires  $U={\rm diag}(u,u)$. 
On $\ga$ and $\sigma$, this translates into:
\debut
\ga \to u\, \ga\, u^{\dag}\quad ;\quad \sigma \to u\,\sigma\,u^T
\label{modulo}
\fin 
with $u$ unitary.
The unitarity and the order two constraints on $P$ and $C$ imply:
\debut
\ga\ga^{\dag}=1,\ \ga^2=1\quad ;\quad \sigma\sigma^{\dag}=1,\
\sigma^T=\pm\sigma
\label{qsigma}
\fin
These conditions are covariant under the 
 gauge transformations (\ref{modulo}).

Let us first only impose a type $P$ symmetry.
Modulo (\ref{modulo}) we can reduce $\ga$ to a diagonal matrix 
with only $\pm 1$ on the diagonal. We may thus  choose:
\debut
        {\rm case\ 1)}\quad &:&\quad  \ga = 1 \label{case1}\\
        {\rm case\ 2)}\quad &:&\quad  \ga =  \sigma_z \otimes 1 \label{case2}
\fin        
In the second case, we assumed for simplicity that the
numbers of $+1$ and $-1$ in $\ga$ are equal, but this  could
 be generalized.

Let us now  impose only  a type $C$ symmetry.
Up to the transformations (\ref{modulo}),
there are two (standard) cases \cite{Dyson} 
depending on the condition $\sigma^T=\pm \sigma$:
\debut
        {\rm case\ 3)}\quad &:&\quad \sigma = 1 \label{case3}\\
        {\rm case\ 4)}\quad &:&\quad \sigma = i\sigma_y\otimes 1 \label{case4}
\fin
Indeed, assume that $\sigma^T=\sigma$. Then, since $\sigma$ is also unitary,
$\sigma\,\sigma^*=1$ and its real and imaginary parts commute and
are both symmetric. They can be simultaneously  diagonalized by a
real orthogonal matrix $o$, so that $\sigma= o \delta o^T$ with
$\delta$ a diagonal unitary matrix. Hence, $\sigma= u\, u^T$,
with $u= o\, \delta^{1/2}$ unitary,
and $\sigma$ is equivalent to the identity modulo (\ref{modulo}). 
The argument is similar for $\sigma^T=-\sigma$.

Next we impose simultaneously a type $P$ and a type $C$ symmetries.
These symmetries  have to be compatible in the sense that their actions
on hamiltonians should commute. For generic hamiltonians this
requires that $C\propto PCP^T$, or $\sigma \propto \ga \sigma \ga^T$.
This condition is covariant under transformations (\ref{modulo}).
So we may choose a basis in which $\ga$ is diagonal and
we restrict ourselves to the two cases (\ref{case1},\ref{case2}).
We then  have two sub-cases corresponding to the two
possible values $\pm 1$ of the  proportionality coefficient 
in the above equation, so that $\sigma$ either commutes or anticommutes with $\ga$:
\debut
 \sigma=\pm\,  \ga\, \sigma\, \ga^T\quad \Rightarrow\quad
[\sigma\, ,\, \ga ] = 0  \quad{\rm or}\quad  \{\sigma\, ,\,  \ga \} =0
\label{qsig}
\fin
If $\ga=1$, $\sigma$ automatically commutes with it and we obtain:
\debut
  {\rm case\ 5)}\quad &:& \quad   \ga = 1,\ \sigma = 1 \label{case5}\\
  {\rm case\ 6)}\quad &:& \quad  \ga = 1,\ \sigma = i\sigma_y \otimes 1 \label{case6}
\fin        
If $\ga=\sigma_z \otimes 1$, we have to consider separately
the two possibilities in (\ref{qsig}). The transformations
(\ref{modulo}) have to preserve the form of $\ga$ so that
$u$ has to be block diagonal $u={\rm diag}(u_1,u_2)$
with $u_{1,2}$ unitary.
When $[\sigma,\ga]=0$, $\sigma$ has also to be block diagonal,
$\sigma={\rm diag}(\sigma_1,\sigma_2)$. As above, modulo (\ref{modulo})
with $u={\rm diag}(u_1,u_2)$, it can be reduced to 
$\sigma=1$ if $\sigma^T=\sigma$ and $\sigma=i\sigma_y\otimes 1$
if $\sigma^T=-\sigma$. Thus, we get two possibilities:
\debut
{\rm case\ 7)}\quad &:& \quad \ga = \sigma_z \otimes 1 ,\ \sigma=1_2\otimes 1
\label{case7}\\
{\rm case\ 8)}\quad &:& \quad \ga = \sigma_z \otimes 1_2\otimes 1 ,\ 
\sigma=1_2\otimes i\sigma_y \otimes 1 \label{case8}
\fin
When $\{\sigma,q\}=0$, $\sigma$ has to be block off-diagonal,
so that $\sigma=\pmatrix{0&s\cr \pm s^T&0\cr}$, with $s$ unitary,
depending whether  $\sigma$ is symmetric or antisymmetric.
The gauge transformations (\ref{modulo}) then become
$s\to u_1\,s\,u_2^T$ with $u_{1,2}$ unitary,
and any unitary $s$ is gauge equivalent to the identity.
This gives two  cases:
\debut
{\rm case\ 9)}\quad &:& \quad \ga = \sigma_z \otimes 1 ,\ \sigma=i\sigma_y\otimes 1
\label{case9}\\
{\rm case\ 9')}\quad &:& \quad \ga = \sigma_z \otimes 1 ,\ \sigma=\sigma_x\otimes 1
\non
\fin
Cases $9)$ and $9')$ turn out to be
equivalent because the type $C$ symmetry of one
of the two cases follows from the product of the type $P$ and the
type $C$ symmetries of the other case.

Finally let us consider more combinations of type $P$ or $C$ symmetries.
If we impose two symmetries of type $P$, their product commutes
with the hamiltonians and this system thus does not correspond 
to a minimal class. 
Next consider imposing  two compatible symmetries of type
$C$ with sign $\ep_{c1}$ and $\ep_{c2}$.
If  the product $\ep_{c1}\ep_{c2}=-1$, their product 
(see eq. (\ref{PC}) )  makes a type $P$
symmetry. Thus two type $C$ symmetries with opposite $\ep_c$ signs
are equivalent to a type  $P$ and a type $C$ symmetries which
we have already classified.  
If the $\ep_c$ signs are equal,
the product of the two type  $C$ symmetries commutes with the hamiltonians
and this system is not minimal.
More generally, considering more combinations of type $P$ and type $C$ symmetries
does not lead to new minimal classes.

\subsection{List of classes.}

In this sub-section we present the detailed structure of the resulting
classes of hamiltonians.  
The type $P$ and $C$ symmetries impose the following relations on
the generalized potentials and masses:
\debut
{\rm P\ sym.}&:&\quad
\ga\,A_z= A_z\, \ga \quad,\quad \ga\,A_\zb= A_\zb\,\ga
\quad,\quad \ga\, V_\pm + V_\pm\, \ga =0 \label{Prel}\\
{\rm C\ sym.}&:&\quad
\sigma\, A_z^T + A_z\, \sigma =0  \quad,\quad 
\sigma\, A_\zb^T + A_\zb\, \sigma =0 \quad,\quad 
\sigma\, V_{\pm}^T = \pm\ep_c\, V_{\pm}\, \sigma \label{Crel}
\fin
They possess a simple interpretation as they indicate that
$A_z$ and $A_\zb$ belong to an orthogonal or symplectic
Lie algebra depending whether $\sigma$ is symmetric or
antisymmetric.
The compatibility relations (\ref{qsig}), $\sigma=\pm \ga\, \sigma\, \ga^T$,
ensures that the constraints (\ref{Prel}) and
(\ref{Crel}) may be imposed simultaneously.
When imposing both a  type $P$ and type $C$ symmetry, one 
generates all relations obtained by successive applications of
these symmetries; i.e. all relations associated to elements
of the group generated by the type $P$ and $C$ symmetries are imposed.
As a consequence, there could be different presentations of the same
class depending which generators of this group one selects.
For example, given a type $P$ and a type $C$ symmetry with a sign
$\ep_c$, their product is again a type $C$ symmetry but with
an opposite sign $-\ep_c$. (See eq. (\ref{PC}).) 

\def\skp{ &~&\non\\}

Solutions of the constraints (\ref{Prel},\ref{Crel})
for the set of compatible $\ga$ and $\sigma$ give the
following minimal classes:
\debut
{\rm class\ {\bf 0}~~} &:&\quad A_\zb,\ A_z \in\ gl(n);\ V_\pm\in\ gl(n). 
\label{class0}\\
\skp
{\rm class\ {\bf 1}~~} &:&\quad A_z\in gl(n);\ V_\pm=0. 
\label{class1}\\
\skp
{\rm class\ {\bf 2}~~} &:&\quad A_z={\rm diag}(a_+,a_-),\ a_\pm\in
gl(n);\non\\
&~&\quad V_\pm=\pmatrix{0&v_\pm\cr w_\pm&0\cr},\ v_\pm,\ w_\pm\in gl(n).
\label{class2}\\
\skp
{\rm class\ {\bf 3}_{\ep_c}} &:&\quad 
A_z=\pmatrix{a & b\cr c & d\cr}= -A_z^T \in\ so(n),\ a=-a^T,\, b=-c^T,d=-d^T; \non\\
&~&\quad V_{-\ep_c}=-V^T_{-\ep_c} \in\ so(n);\
V_{\ep_c}=V^T_{\ep_c} \in\ gl(n)\setminus so(n).
\label{class3}\\
\skp
{\rm class\ {\bf 4}_{\ep_c}}  &:&\quad A_z=\pmatrix{a & b\cr c & d\cr}
=-\sigma_y A_z^T\sigma_y \in\ sp(2n),\ a=-d^T,\, b=b^T,c=c^T; \non\\
&~&\quad V_{-\ep_c}=-\sigma_y V_{-\ep_c}^T\sigma_y \in\ sp(2n);\
V_{\ep_c}=\sigma_y V_{\ep_c}^T\sigma_y \in\ gl(2n)\setminus sp(2n).
\label{class4}\\
\skp
{\rm class\ {\bf 5}~~}  &:&\quad A_z=-A_z^T \in\ so(n);\ V_\pm=0.
\label{class5}\\
\skp
{\rm class\ {\bf 6}~~}  &:&\quad A_z=-\sigma_y A_z^T\sigma_y \in\ sp(2n);\ V_\pm=0.
\label{class6}\\
\skp
{\rm class\ {\bf 7}~~} &:& \quad
A_z={\rm diag}(a_+,a_-),\ a_\pm=-a_\pm^T\in so(n);\non\\
 &~& \quad V_{\pm}=\pmatrix{ 0 & v_{\pm}\cr w_{\pm} &0\cr},\
v_{\pm\ep_c}=\pm  w_{\pm\ep_c}^T.
\label{class7}\\
\skp
{\rm class\ {\bf 8}~~} &:& \quad
A_z={\rm diag}(a_+,a_-),\ a_\pm=-\sigma_y a_\pm^T \sigma_y \in sp(2n);\non\\
 &~&\quad V_{\pm}=\pmatrix{ 0 & v_{\pm}\cr w_{\pm} &0\cr},\
v_{\pm\ep_c}=\pm \sigma_y w_{\pm\ep_c}^T \sigma_y.
\label{class8}\\
\skp
{\rm class\ {\bf 9}_{\ep_c}} &:& \quad
A_z={\rm diag}(a,-a^T),\ a \in gl(n);\non\\
 &~&\quad V_{\pm \ep_c}=\pmatrix{ 0 & v_{\pm \ep_c}\cr w_{\pm \ep_c} &0\cr},\
v_{\pm \ep_c}=\mp  v_{\pm \ep_c}^T,\ w_{\pm \ep_c}=\mp  w_{\pm \ep_c}^T
\label{class9}
\fin
(The labels 1-9 refer to the cases 1-9 listed in the previous subsection.) 
The hermiticity constraints $A_\zb=A_z^{\dag}$, $V_\pm=V_\pm^{\dag}$, 
$v_\pm^{\dag}=w_\pm$, are implicit in this list.  (The hermiticity constraint
is not made explicit for the purpose of describing the non-hermitian classes
in the next section.) 
The index $\ep_c$ refers to one of the two possible values $\ep_c=\pm$; and  
the absence of such index means that this value is irrelevant.
The first class corresponds to generic Dirac hamiltonians with no
constraints imposed. There  are a few degeneracies when solving
eqs.(\ref{Prel},\ref{Crel}). As expected, cases $9$ and $9'$,
eq.(\ref{case9}), yield the same solutions; also
 realizations of the cases $5$ and $6$, eqs.(\ref{case5},\ref{case6})
are independent of the choice of the sign $\ep_c$.
Realisations of the cases $7_\pm$  are also equivalent because 
they correspond to different presentations of the same class.
Indeed, consider eq.(\ref{case7}) in case $7$.
The product of its type $P$ and type $C$ defining symmetries produces
a new type $C$ symmetry with opposite sign and with
$\sigma'=\sigma_z\otimes 1$. It is gauge equivalent to 
the original symmetries as $\sigma'\simeq u_7\, \sigma'\,u_7^T =
1_2\otimes 1$ with $u_7={\rm diag}(1,i)\otimes 1$.
Hence case $7_+$ and $7_-$ are gauge equivalent, and
we give the two presentations in the above list.
The corresponding realizations are related by similarity
transformations $H\to H'=UHU^{\dag}$ with 
$U={\rm diag}(u_7,u_7)$, so that $a_\pm'=\pm a_\pm$
and $v'_\pm=-iv_\pm$, $w'_\pm=iw_\pm$. A similar argument applies to the
cases $8_\pm$ showing again that they are equivalent presentations
of the same class.

One of the origins  of the distinction  between the above 
classification and the classification of random matrices
arises from a difference in the notion of
equivalent classes of compatible symmetries.
For random matrices,  some of the cases with  $\ga=1$ or $\ga=\sigma_z\otimes 1$ 
are considered as equivalent because
they correspond to the same operator $P$ up to re-shuffling
of the lines and columns, while in the present classification
they yield different classes because  we impose the
$2\times 2$ block structure (\ref{hamilD}) to the hamiltonians
\footnote{The classification of random matrices may be reread
from the previous classification  by considering that 
$\ga$ and $\sigma$ implement directly the discrete type $P$ and $C$ symmetries.
Only the cases with $\ga=\sigma_z\otimes 1$ is then relevant since $\ga=1$
is trivial. As it should be, we are left with ten classes
${\bf 0},\ {\bf 2},\ {\bf 3}_\pm,\  
{\bf 4}_\pm,\ {\bf 7},\ {\bf 8},\ {\bf 9}_\pm$.
However, in Table 1, the random matrix classes refer to those defined
by $C$ and $P$ and not by $\ga$ and $\sigma$. See the appendix.}.
Thus the classification of random Dirac operators is a bit finer
than the one of random matrices, as summarized in  Table 1.

\subsection{Symmetry groups,  disorder measures, and super-symmetric effective actions.}

Each class is stable under the action of a symmetry group,
whose elements act on the hamiltonians by conjugation  such 
that their form imposed by eqs.(\ref{class0}--\ref{class9}) 
is preserved. Elements $G$ of the symmetry groups satisfy
\footnote{One may extend slightly the symmetry group by discrete groups,
made of ${\bf Z}_2$ factors, by allowing signs in eq.(\ref{groups}), 
$G\,\ga\,G^{-1}=\pm \ga$, $ G\,\sigma\,G^T=\pm\sigma$.}:
\debut
G\,\ga\,G^{-1}=\ga \quad ,\quad 
G\,\sigma\,G^T=\sigma
\label{groups}
\fin
The list of these groups is given in Table 1.
For classes ${\bf 2},\ {\bf 7},\ {\bf 8}$, in which the symmetry
group is a product of two subgroups, the embedding is diagonal
with $G={\rm diag}(g_+,g_-)$ where $g_\pm$ belong either to
$U(n)$, $O(n)$ or $Sp(2n)$.  In class ${\bf 9}$ the embedding
is $G={\rm diag}(g,g^T)$ with $g\in U(n)$.

The symmetry group  may be used to specify the disorder
measures in each class,  which we assume to be
Gaussian, with zero mean, and local.
The measures are then fixed by requiring them to be 
invariant under the symmetry group. 
The list of all quadratic invariants for each class
is the following:

\debut 
{\rm class}\ {\bf 0}~~ &:& tr(A_zA_\zb),\ tr(A_z)tr(A_\zb),\ tr(V_\pm^2),\ 
tr(V_\pm)^2,\ tr(V_\pm V_\mp),\ 
tr(V_\pm)tr(V_\mp); \non\\
{\rm class}\  {\bf 1}~~ &:&  tr(A_zA_\zb),\ tr(A_z)tr(A_\zb);\non \\
{\rm class}\  {\bf 2}~~ &:& tr(a_\pm\bar a_\pm),\ tr(a_\pm)tr(\bar a_\pm),\ 
tr(v_\pm^{\dag}v_\pm),\ tr(v_\pm^{\dag}v_\mp) ;\non\\ 
{\rm class}\  {\bf 3}_{\ep_c} &:&  tr(A_zA_\zb),\ tr(V_{\pm\ep_c}^2),\ 
tr(V_{\ep_c})^2;  \non\\ 
{\rm class}\  {\bf 4}_{\ep_c} &:&  tr(A_zA_\zb),\ tr(V_{\pm\ep_c}^2), \
tr(V_{\ep_c})^2;  \non\\ 
{\rm class}\  {\bf 5}~~ &:& tr(A_zA_\zb);\non\\ 
{\rm class}\ {\bf 6}~~ &:& tr(A_zA_\zb);\non\\ 
{\rm class}\ {\bf 7}~~ &:&  tr(a_\pm\bar a_\pm),\ 
tr(v_\pm^{\dag}v_\pm);\non \\
{\rm class}\  {\bf 8}~~ &:&  tr(a_\pm\bar a_\pm),\ 
tr(v_\pm^{\dag}v_\pm);\non\\ 
{\rm class}\  {\bf 9}_{\ep_c} &:& tr(a\bar a),\ tr(a)tr(\bar a),\
tr(v_\pm^{\dag}v_\pm),\ tr(v_\pm^{\dag}v_\mp) .\non
\fin  
To preserve rotation invariance we only list the invariants 
which couple $A_z$ to $A_\zb$ and $V_\pm$ to itself or to $V_\mp$.
Couplings between  $V_\pm$ and $V_\mp$ break the  symmetry
under reflection $x\to x,\ y\to -y$
 
These ensembles may be analyzed using the supersymmetric method
\cite{Efetov}. For  each class, this leads to an effective  field theory description
with the  number of coupling constants equal to the number of invariants.
These coupling constants, which measure the strength
of the disorder, parameterize perturbations of the
free field theory valid in the absence of disorder.
In two dimensions, 
all of these effective field theories can be formulated as left-right
current-current perturbations, where the couplings are marginal\cite{Cargese}.   
This  means that in the effective
field theory the coupling constants are dual
to operators of scaling dimension two.
The discrete symmetry defining the classes plus the global
invariance under the symmetry group should ensure that for  each class
the effective field theory is perturbatively
renormalizable, i.e. no additional  marginal operators beyond  
those dual  to the coupling constants are 
generated by the renormalization procedure.
This aspect can be studied using the all-orders $\beta$ function proposed in
\cite{GLM}, as was done for class {\bf 0}   and 
for class {\bf 4}$_-$ at $n=1$\cite{net}.      

We can easily describe the global Lie superalgebra symmetry of the effective 
field theories.  The unperturbed conformal field theory has an $osp(2N|2N)$ 
current algebra symmetry at level 1 where $N$ is the number of
fermions, i.e. $N=n$ for classes {\bf 0, 1, 3} and {\bf 5},  
$N=2n$ for classes {\bf 2, 4, 6, 7} and {\bf 9}, and $N=4n$ for class 
{\bf 8}.  
In the supersymmetric effective theory the global supersymmetry is smaller,
and corresponds to the Lie superalgebraic extension $\CG$ of the symmetry groups listed
in Table 1.    The bosonic group $U(n)$ is extended to $gl(n|n)$,  $O(n)$ to 
$osp(n|n)$ and $sp(2n)$ to $osp(2n|2n)$, so that $\CG = gl(n|n)$ for classes 
{\bf 0,1,9},
$\CG=osp(n|n)$ for classes {\bf 3,5},  
$\CG = osp(2n|2n)$ for classes {\bf 4,6}  and a tensor
product of these supergroups for classes {\bf 2,7,8}.

\section{Classification of non-hermitian Dirac hamiltonians.}

The classification of non-hermitian Dirac hamiltonians we present
is based on implementing  discrete symmetries of type $P$ or $C$,
eqs.(\ref{Psym},\ref{Csym}), and  of type $Q$ and $K$ defined in
eqs.(\ref{Qsym},\ref{Ksym}). Since this parallels closely what we have done  
for  hermitian hamiltonians we only sketch the main features of the classification.  
The order two constraints on type $Q$ and $K$ symmetry are:
\debut
\xi\xi^{\dag}=1,\ \xi^2=1 \quad;\quad
\kappa\kappa^{\dag}=1,\ \kappa^T=\pm \kappa \non
\fin
We consider hamiltonians up to unitary
changes of basis, $H\to UHU^{\dag}$, with $U={\rm diag}(u,u)$, 
which act on $Q$ and $K$ as:
\debut
\xi\to u\,\xi\,u^{\dag} \quad;\quad 
\kappa\to u\,\kappa\,u^T
\label{mod2}
\fin
As before we define minimal classes as  those whose hamiltonians do not
commute with a fixed matrix preserving their Dirac structure.

Imposing the type $Q$ or $K$ symmetries amounts
to imposing  some reality conditions on  the hamiltonians, i.e.
some reality properties of $A_z,\ A_\zb$ and $V_\pm$:
\debut
{\rm Q\ sym.}&:&\quad
\xi\,A_\zb^{\dag}= A_z\,\xi
\quad,\quad \xi\, V_\pm^{\dag} =\ep_q\, V_\pm\, \xi \label{Qrel}\\
{\rm K\ sym.}&:&\quad
\kappa\, A_\zb^* + A_z\, \kappa =0 \quad,\quad 
\kappa\, V_{\pm}^* = \pm\ep_k\, V_{\pm}\, \kappa \label{Krel}
\fin
Redefining $H\to iH$ modifies the signs $\ep_q$ and $\ep_k$ in
eqs.(\ref{Qsym},\ref{Ksym}), however  this redefinition ruins  
the Dirac structure (\ref{hamilD}) and we  shall thus not allow it.

The classification of non-hermitian hamiltonians  may be translated into detailed properties
of the hermitian hamiltonians $\cal H$ obtained by doubling 
the Hilbert spaces on which the Dirac hamiltonians $H$ are acting:
\debut
\CH = \pmatrix{ 0 & H\cr H^{\dag} & 0\cr}
\label{double}
\fin
These doubled hamiltonians are always chiral as they anticommute
with $\Ga_5={\rm diag}(1,-1)$. 
Any similarity transformation $H\to UHU^{-1}$ is mapped into
$\CH\to\CU\CH\CU^{\dag}$ with $\CU={\rm diag}(U,U^{\dag\,-1})$.
Demanding that these transformations also act by similarity
on $\CH$ imposes $U$ to be unitary.
When no discrete symmetries are imposed, 
the doubled hamiltonians $\CH$ are always elements of class ${\bf 2}$,
which is embedded in the chiral GUE class. Indeed, up to
re-shuffling of lines and columns, they may be presented as:
\debut
\CH \simeq \CH_d\equiv
\pmatrix{ 0 & V_+^{\dag} + V_-^{\dag} & -i\d_\zb + A_z^{\dag} & 0 \cr
    V_+ + V_- & 0 & 0 & -i\d_\zb + A_\zb \cr
   -i\d_z + A_z & 0 & 0 & V_+ - V_- \cr
   0 & -i\d_z + A_\zb^{\dag} & V_+^{\dag} - V_-^{\dag} & 0 \cr}
\label{chdouble}
\fin
The dictionary is thus $a_+=A_z$, $a_-= A_\zb^{\dag}$
and $2v_\pm= (V_+^{\dag}\pm V_+)+(V_-^{\dag}\mp V_-)$.

On $\CH$, both type $P$ and $Q$ symmetries act as chiral
transformations, $\CH\to -\CP\CH\CP^{-1}$ with $\CP={\rm diag}(P,P)$
and $\CH\to \ep_q\CQ\CH\CQ^{-1}$ with $\CQ=\pmatrix{0&Q\cr Q&0\cr}$.
Thus, $\CH$ may be block diagonalized if $H$ is $Q$ or $P$ symmetric.
Indeed, if $H$ is $Q$ symmetric with $\ep_q=+1$
then $\CQ$ and $\CH$ may be simultaneously diagonalized 
since they commute. Similarly, if $H$ is $P$ or $Q$ symmetric 
with $\ep_q=-1$, then $\CH$ commutes with the product $\Ga_5\CP$ or 
with $\Ga_5\CQ$. 

Type $C$ and $K$ symmetries both act as particle-hole
symmetries relating $\CH$ to its transposed $\CH^T$.
The classification of the hamiltonians $\CH$ is then very
simple as it follows from that of hermitian Dirac operators.
For the operators $\CH$ to be minimal, 
 only a type $C$ or a type  $K$ symmetry can  be imposed. 
Gauge equivalences (\ref{modulo},\ref{mod2}) leave only
$\sigma=1$ or $\sigma=i\sigma_y\otimes1$ and
$\kappa=1$ or $\kappa=i\sigma_y\otimes 1$ as possible choices.
We then have the correspondence:
\debut
\sigma=1 &\Rightarrow & \CH\in\ {\rm class}\ {\bf 7} \label{CHclass}\\
\sigma=i\sigma_y\otimes1 &\Rightarrow & \CH\in\ {\rm class}\ {\bf 8} \non\\
\kappa=1\ {\rm or}\ \kappa=i\sigma_y\otimes1
 &\Rightarrow & \CH\in\ {\rm class}\ {\bf 9}_\pm \non
\fin

Though we have translated the classification of non-hermitian $H$ into
the doubled hermitian $\CH$,  the spectrum of $\CH$ and $H$ may differ significantly.
To illustrate these potential differences, consider the transformation
$H\to \tilde H = -i u_7\, H\, u_7$ where $u_7={\rm diag}(1,i)\otimes1$
is the matrix we introduced in Section II. This transformation
preserved the Dirac structure of the hamiltonians but not
their reality conditions. It leaves invariant $A_z$ and $A_\zb$ but
not the potentials since $\tilde V_\pm=-iV_\mp$.
Hence $H$ and $\tilde H$ should not belong
to the same non hermitian class --- and they do not have the same spectra.
On the  contrary, for the doubled hamiltonians this transformation is lifted to
$\CH\to \tilde \CH=\CU\,\CH\,\CU^{\dag}$ with
 $\CU={\rm diag}(-iu_7,u_7^{\dag})$, so that $\CH$ and $\tilde \CH$
are unitarily equivalent. They have identical spectra and
belong to the same class.

We thus present a classification of the Dirac operators
$H$ and not simply of the doubled one.
As for hermitian Dirac operators, it is the group generated 
by compatible discrete symmetries which is meaningful. 
There  could be different but equivalent presentations
of the same group as not all of these symmetries are independent.
Indeed, the product of a type $P$ symmetry with  a $C, K$ or $Q$ symmetry
is again a $C,K$ or $Q$ symmetry.  (See for example eq. (\ref{PC}).)  
Also, the symmetries of
type $Q$, $C$ or $K$ are linked as the product of any of two of them
gives a symmetry of the third type. 

Let us first impose only one type of symmetry.
As in Section II, up to gauge equivalence 
(\ref{modulo},\ref{mod2}), the solutions are:
\begin{center}
\begin{tabular}{cccc}
$(\ga=1)\quad ;$ & $(\sigma=1)_{\ep_c}\quad ;$ 
& $(\xi=1)_{\ep_q}\quad ;$ & $(\kappa=1)_{\ep_k}\quad ;$ \\
$(\ga=\sigma_z\otimes1)\quad ;$ & $(\sigma=i\sigma_y\otimes 1)_{\ep_c}\quad ;$ 
& $(\xi=\sigma_z\otimes 1)_{\ep_q}\quad ;$ 
&  $(\kappa =i\sigma_y\otimes 1)_{\ep_k}\quad ;$ \\
\end{tabular}
\end{center}
Here, 
each column refers to one of the possible types of symmetry.
Here and below, we indicate as indices the values
of $\ep_c$, $\ep_q$ or $\ep_k$ which matter.
Thus the above list corresponds to $14$ distinct classes.  

Let us now impose two kinds of symmetry.
First, we may require simultaneously type $P$ and $C$ symmetries.
This leads to the list of 6  classes, from
class ${\bf 5}$ to  class ${\bf 9}$ of
Section II without any reality conditions.

Next we consider a $P$ and a $K$ symmetry. 
The commutativity condition for type $P$ and $K$ symmetries
reads $\kappa=\pm \ga^{-1}\kappa\,\ga^*$. This is solved
the same way as $\sigma=\pm \ga^{-1}\sigma\ga^T$ in previous
Section. Thus the list of compatible type $P$ and $K$ symmetries
is parallel to the list of compatible type $P$ and $C$ symmetries,
only $\sigma$ is replaced by $\kappa$.
Their explicit realizations are given in
eqs.(\ref{class5}--\ref{class9}) but with $v_\pm^T$ 
replaced by $w_\pm^*$ and $w_\pm^T$ replaced by $v_\pm^*$.  
This corresponds to six  classes.  

Similarly compatibility between type $P$ and $Q$ symmetries
requires $\ga^{\dag}=\pm \xi^{-1}\ga\,\xi$.
Solving this constraint leads to the following compatible
type $P$ and $Q$ symmetries:
\debut
&& (\ga=1,\ \xi=1) \ ;\quad (\ga=1\otimes 1,\ \xi=\sigma_z\otimes1)\ ;
\quad (\ga=\sigma_z\otimes 1,\ \xi=\sigma_x\otimes 1)_{\ep_q}  ;\non\\
&& ~~~~~~~~~   (\ga=\sigma_z\otimes 1,\ \xi=1\otimes 1)_{\ep_q}
\cong (\ga=\sigma_z\otimes 1,\ \xi=\sigma_z\otimes1)_{-\ep_q}\ ; \non
\fin
In the second line, we have mentioned an equivalence between 
two solutions of the commutativity constraint. Indeed, $\xi$
of the second solution in this line is the product of
$\ga$ and $\xi$ of the first solution, so the groups generated
by these solutions are identical.

We may also impose together a type $Q$ with a type $C$
symmetry. Since their product is a symmetry of type $K$ with
$\ep_k=\ep_q\ep_c$, we are actually imposing simultaneously
three compatible symmetries of different types. 
Any two of them generate the third.
The condition for the type $Q$ and $C$ symmetries to commute
is $\xi^T=\pm \sigma^{\dag}\, \xi^{-1}\, \sigma$,
for type $C$ and $K$ this condition reads 
$\kappa^T\,\sigma^{-1}\,\kappa\, \sigma^*=\pm1$. 
Up to gauge equivalences, the set of compatible
type $Q$ and $C$ symmetries is then:
\begin{center}
\begin{tabular}{ccc} 
$(\xi=1,\ \sigma=1)_{\ep_q,\ep_c} ;$ 
& $(\xi=\sigma_z\otimes1,\ \sigma=1\otimes1)_{\ep_q,\ep_c} ;$
& $(\xi =\sigma_z\otimes 1,\ \sigma=i\sigma_y\otimes 1)_{\ep_q,\ep_c} ;$ \\
$(\xi=1\otimes1,\ \sigma=i\sigma_y\otimes1)_{\ep_q,\ep_c} ;$ 
& $(\xi=\sigma_z\otimes1,\ \sigma=1\otimes i\sigma_y)_{\ep_q,\ep_c} ;$
& $(\xi =\sigma_z\otimes 1,\ \sigma=\sigma_x\otimes 1)_{\ep_q,\ep_c}\;$\\
\end{tabular}
\end{center}

Finally, we may impose simultaneously a type $P$ symmetry
together with two among the three types $Q$, $C$ and $K$ of symmetries.
As before it is sufficient to consider only a $P, C$ and $Q$ symmetry.  
The solutions of the commutativity requirements are then:
\debut
&&(\ga=1\otimes1,\ \xi =1\otimes1,\ 
\sigma=1\otimes1\ {\rm or}\ i\sigma_y\otimes 1) ; \non\\
&&(\ga=1\otimes1,\ \xi =\sigma_z\otimes1,\  
\sigma=1\otimes1\ ,  \sigma=1\otimes i\sigma_y\
,  i\sigma_y\otimes 1\  {\rm or}\
 \sigma_x\otimes 1) ;\non\\
&&(\ga=\sigma_z\otimes1,\ \xi =1\otimes1,\ 
\sigma=1\otimes1\ {\rm or}\ 1\otimes i\sigma_y )_{\ep_q} ; \non\\ 
&&(\ga=\sigma_z\otimes1,\ \xi =1\otimes1,\ 
\sigma=\sigma_x \otimes 1 )_{\ep_q, \ep_c}  ;\non\\
&&(\ga = \sigma_z \otimes 1, \xi=\sigma_x \otimes 1, 
\sigma= 1\otimes 1,  1\otimes i\sigma_y, \sigma_x \otimes 1,\ {\rm or} \  
\sigma_x \otimes i\sigma_y \otimes 1  )_{\ep_q, \ep_c} 
\non
\fin
Here, each choice of $\sigma$ corresponds to a different class.

Considering more combinations of the four different kinds of
symmetries would not lead to new minimal classes, because in such case we
would always be able to construct some matrix commuting with the
hamiltonians and preserving their Dirac structure.

For each set of compatible symmetries, one has to choose the signs
$\ep_q$, $\ep_c$ and $\ep_k$ to specify the classes. These
signs are used to index the solutions in the above lists.
The absence of one of this index means that the corresponding
solution is independent of that index.
The grand total is $87$ classes.  

It is straightforward to determine the form of $H$ for each of the above symmetry
classes,  however there is little motivation to list the details here.  
Let us just describe a simple example,   
corresponding  to imposing  a symmetry of type $K$, relating
$H$ to its complex conjugate, with $\kappa=1$ and $\ep_k=+1$.
Then, relations (\ref{Krel}) yield $A_\zb=-A_z^*$ and 
$V_\pm=\pm V^*_\pm$, such that $V_+$ is real and $V_-$ imaginary.
As a consequence, the Dirac hamiltonian may be written as:
\debut
 \left(\matrix{ M  & -i \d_\zb - A_z^* 
         \cr  -i \d_z + A_z & M^* \cr} \right) \non
\fin
with $M=V_++V_-$. This class, studied e.g. in ref.\cite{GLL},
is closely related to the random $XY$ model.
The doubled hamiltonian belongs to class ${\bf 9}_-$.

Having performed the above classification we can easily specialize
it to random non-hermitian matrices with no Dirac structure. 
As for random hermitian matrices (see the appendix for a summary),  
the above classes with
$\gamma = 1$  are trivial and should be thrown away.   
The choice of the sign $\ep_q$ and $\ep_k$  
is also irrelevant since  it can be   absorbed into $H \to iH$ 
which is  now allowed since no Dirac  structure is imposed.
Altogether this  gives 43 classes which will be described in 
greater detail in \cite{tocome}.

%
%

\section{Appendix.}

For completeness --- and for explaining Table 1 ---
we recall the definition of random hermitian matrix ensembles \cite{Zirn}.
We denote by small letters quantities referring to random matrices.
Let $h=h^{\dag}$ be a hermitian matrices and $p$ and $c$ be
the operators implementing the discrete symmetries as 
in eqs.(\ref{Prel},\ref{Crel}): $h\to -php^{-1}$,
$h\to \ep_c\, ch^Tc^{-1}$.
For each random matrix class, the defining relations 
for $p$ and $c$ are summarized in Table 2.

\begin{center}
\begin{tabular}{|c|c|}
\hline
 Random matrix & Discrete sym.  \\
   classes      &  Relations  \\
\hline
\hline
${\bf A}$~~ & $h=h^{\dag}$ \\
\hline
${\bf A}_I$ & $c^T=c,\ \ep_c=+$ \\
\hline
${\bf A}_{II}$ & $c^T=-c,\ \ep_c=+$ \\
\hline
${\bf A}_{III}$ & $p^2=1$ \\
\hline
${\bf C}$~~ & $c^T=-c,\ \ep_c=-$ \\
\hline
${\bf D}$~~ & $c^T=c,\ \ep_c=-$ \\
\hline
${\bf D}_{I}$ & $p^2=1,\ c^T=c,\ \ep_c=\pm,\ pcp^T=c$\\
\hline
${\bf C}_{II}$ & $p^2=1,\ c^T=-c,\ \ep_c=\pm,\ pcp^T=c$\\
\hline
${\bf C}_{I}$ & $p^2=1,\ c^T=\pm c,\ \ep_c=\pm,\ pcp^T=-c$\\
\hline
${\bf D}_{III}$ & $p^2=1,\ c^T=\pm c,\ \ep_c=\mp,\ pcp^T=-c$\\
\hline
\end{tabular}
\end{center}

In each of the last four lines of Table 2, one may 
equivalently choose either the upper or the lower signs,
since this simply corresponds
to choosing  two equivalent presentations of the same class.

To compare with the classification of Dirac fermions, it
is  useful to notice that in the latter case
the operators $P$ and $C$ may be written as:
\debut
P=\tau_z\otimes \ga;\quad
C=i\tau_y\otimes \sigma\ {\rm for}\ \ep_c=+;\quad
C=\tau_x\otimes \sigma\ {\rm for}\ \ep_c=-. \non
\fin

\end{document}